\documentclass[10pt,twocolumn,english,journal]{IEEEtran}
\usepackage[T1]{fontenc}
\usepackage[latin9]{inputenc}
\usepackage{amsmath}
\usepackage{graphicx}
\usepackage{esint}

\makeatletter
\newtheorem{definitn}{Definition}
\newtheorem{lemma}{Lemma}
\newtheorem{thm}{Theorem}

\usepackage{amssymb,amsmath}

\usepackage{babel}
\makeatother

\begin{document}

\title{Scaling Laws for Overlaid Wireless Networks: A Cognitive Radio Network
vs. a Primary Network}

\author{Changchuan Yin, Long Gao, and Shuguang Cui%
\thanks{The authors are with the Department of Electrical and Computer Engineering,
Texas A\&M University, TX 77843, USA (e-mail: ccyin@ieee.org, lgao@ece.tamu.edu,
cui@tamu.edu).%
}}

\maketitle
\begin{abstract}
We study the scaling laws for the throughputs and delays of two coexisting
wireless networks that operate in the same geographic region. The
primary network consists of Poisson distributed legacy users of density
$n$, and the secondary network consists of Poisson distributed cognitive
users of density $m$, with $m>n$. The primary users have a higher
priority to access the spectrum without particular considerations
for the secondary users, while the secondary users have to act conservatively
in order to limit the interference to the primary users. With a practical
assumption that the secondary users only know the locations of the
primary transmitters (not the primary receivers), we first show that
both networks can achieve the same throughput scaling law as what
Gupta and Kumar~\cite{Capacity:Gupta} established for a stand-alone
wireless network if proper transmission schemes are deployed, where
a certain throughput is achievable for each individual secondary user
(i.e., zero outage) with high probability. By using a fluid model,
we also show that both networks can achieve the same delay-throughput
tradeoff as the optimal one established by El Gamal \emph{et al.}~\cite{Delay:Gamal1}
for a stand-alone wireless network. 
\end{abstract}

\section{Introduction}

Initiated by the seminal work of Gupta and Kumar~\cite{Capacity:Gupta},
the throughput scaling law for large-scale wireless networks has become
an active research topic~\cite{Gap:Fran}-\cite{Hierarchical:Ozgur}.
Scaling laws provide a fundamental way to measure the achievable throughput
of a wireless network. Considering $n$ nodes that are randomly distributed
in a unit area and grouped independently into one-to-one source-destination
(S-D) pairs, Gupta and Kumar~\cite{Capacity:Gupta} showed that typical
time-slotted multi-hop architectures with a common transmission range
and adjacent-neighbor communication can achieve a sum throughput that
scales as $\Theta\left(\sqrt{n/\log n}\right)$. They also showed
that an alternative arbitrary network structure with optimally chosen
traffic patterns, node locations, and transmission ranges can achieve
a sum throughput of order $\Theta\left(\sqrt{n}\right)$. Thus, they
suggested that a factor of $\sqrt{\log n}$ is the price to pay for
the randomness of the node locations. In~\cite{Gap:Fran}, with percolation
theory, Franceschetti \emph{et~al.} showed that the $\Theta\left(\sqrt{n}\right)$
sum throughput scaling is achievable even for randomly deployed networks
under certain special conditions. In~\cite{Mobility:Grosslauser},
Grossglauser and Tse showed that by allowing the nodes to move independently
and uniformly, a constant throughput scaling $\Theta(1)$ per S-D
pair can be achieved. Later, Diggavi \emph{et~al. }showed that a
constant throughput per S-D pair is achievable even with a one-dimensional
mobility model~\cite{Oned:Diggavi}. In these approaches, the network
area is fixed and the throughput scales with the node density $n$.
We call this kind of network as \emph{dense network}. On the other
hand, based on the \emph{extended network} model where the density
of nodes is fixed and the network area increases with $n$, the information-theoretic
scaling laws of transport capacity were studied for different values
of the pathloss exponent $\alpha$ in~\cite{Scaling:Xie}- \cite{Hierarchical:Ozgur}.
In particular, \"Ozg\"ur \emph{et~al.}~\cite{Hierarchical:Ozgur}
proposed a hierarchical cooperation scheme to achieve a sum throughput
that scales as $n^{2-\alpha/2}$ for $2\leq\alpha<3$, i.e., asymptotically
linear for $\alpha=2$. 

In wireless networks, another key performance metric is delay, which
incurs the interesting problems regarding the interactions between
throughput and delay. The issues of delay-throughput tradeoff for
static and mobile wireless networks have been addressed in~\cite{Delay:Gamal1},
\cite{Delay:Toumpis}-\cite{Delay:Ozgur}\emph{.} In~\cite{Delay:Gamal1},
El Gamal \emph{et~al}. established the optimal delay-throughput tradeoff
for static and mobile wireless networks. For static networks, they
showed that the optimal delay-throughput tradeoff is given by $D(n)=\Theta\left(n\lambda(n)\right)$,
where $\lambda(n)$ and $D(n)$ are the throughput and delay per S-D
pair, respectively. Using a random-walk mobility model, they showed
that a much higher delay of $\Theta\left(n\log n\right)$ is associated
with the higher throughput of $\Theta(1)$ for mobile networks. The
delay-throughput tradeoffs in mobile wireless networks have been investigated
under many other mobility models, which include the i.i.d. model~\cite{Delay:Toumpis},
\cite{Delay:Neely}, \cite{Delay:Lin}, the hybrid random walk model~\cite{Delay:Sharma},
and the Brownian motion model~\cite{Delay:Lin1}. For the hierarchical
cooperation scheme in a static wireless network, \"Ozg\"ur and L\'ev\^eque~\cite{Delay:Ozgur}
showed that a significantly larger delay was introduced compared with
the traditional multi-hop scheme, and the delay-throughput tradeoff
is $D(n)=\Theta\left(n\left(\log n\right)^{2}\lambda(n)\right)$ for
$\lambda(n)$ between $\Theta\left(1/(\sqrt{n}\log n)\right)$ and
$\Theta\left(1/\log n\right)$. 

All the aforementioned results focus on the throughput scaling laws
or the delay-throughput tradeoffs for a single wireless network. In
recent years, the ever-growing demand for frequency resource from
wireless communication industries imposes more stress over the already-crowded
radio spectrum. However, a recent report by the Federal Communications
Commission (FCC) Spectrum Policy Task Force indicated that over 90
percent of the licensed spectrum remains idle at a given time and
location~\cite{Federal:Spectrum}. This motivated the regulation
bodies to consider the possibility of permitting secondary networks
to coexist with licensed primary networks, which is the main driving
force behind the cognitive radio technology~\cite{Cognitive:Haykin}.
In a secondary network, the cognitive users opportunistically access
the spectrum licensed to primary users according to the spectrum sensing
result~\cite{Spectrum:Quan}, where the primary users have a higher
priority and the secondary users need to prevent any harmful interference
to the primary users~\cite{Cognitive:Sahai}, \cite{Power:Hoven}.
In this overlaid regime, the throughput scaling law and the delay-throughput
tradeoff for both the primary and secondary networks are interesting
and challenging problems. Some preliminary work along this line appeared
recently. In~\cite{Cognitive:Vu}, \cite{Scaling:Vu}, Vu~\emph{et~al.}
considered the throughput scaling law for a single-hop cognitive radio
network, where a linear scaling law is obtained for the secondary
network with an outage constraint for the primary network. In~\cite{Scaling:Jeon},
Jeon~\emph{et~al}. considered a multi-hop cognitive network on top
of a primary network and assumed that the secondary nodes know the
location of each primary node regardless of whether it is a transmitter
(TX) or a receiver (RX). With an elegant transmission scheme, they
showed that by defining a preservation region around each primary
node, both networks can achieve the same throughput scaling law as
a stand-alone wireless network, while the secondary network may suffer
from a finite outage probability.

In a practical cognitive network, it is hard for the secondary users
to know the locations of primary receiving nodes since they may keep
passive all the time. A reasonable assumption is that the secondary
network knows the locations of the primary TXs. Based on this assumption,
in this paper we define a preservation region just around each primary
TX and propose corresponding transmission schemes for the two networks.
We show that when the secondary network has a higher density as requested
in~\cite{Scaling:Jeon}, both networks can achieve the same throughput
scaling law as a stand-alone wireless network, with zero outage for
the secondary users with high probability. Considering a fluid model,
we also show that both networks can achieve the same delay-throughput
tradeoff as the optimal one established for a stand-alone static wireless
network in~\cite{Delay:Gamal1}. In our approach, the primary network
deploys a time-slotted multi-hop transmission scheme similar to that
in~\cite{Capacity:Gupta} and does not need to cooperate with the
secondary network. Note that, as mentioned in~\cite{Scaling:Jeon},
if both the primary network and the secondary network are willing
to cooperate and do time-sharing, both of them could easily achieve
the same throughput scaling law as a stand-alone wireless network.

The rest of the paper is organized as follows. The system model, definitions,
and main results are described in Section II. The proposed protocols
for the primary and secondary networks are discussed in Section III.
The delay and throughput scaling laws for the primary network are
established in Section IV. The delay and throughput scaling laws for
the secondary network are derived in Section V. Finally, Section VI
summarizes our conclusions.

\section{System Model, Definitions, and Main Results}

In this section, we first describe the system model and assumptions
about the primary and secondary networks, and then define the throughput
and delay. We use $p(E)$ to represent the probability of event $E$
and claim that an event $E_{n}$ occurs with high probability (\emph{w.h.p.})
if $p(E_{n})\to1$ as $n\to\infty$. We use the following order notations
throughout this paper. Given non-negative functions $f(n)$ and $g(n)$:

\begin{enumerate}
\item $f(n)=O(g(n))$ means that there exists a positive constant $c_{1}$
and an integer $m_{1}$ such that $f(n)\leq c_{1}g(n)$ for all $n\geq m_{1}$\emph{.}
\item $f(n)=\Omega(g(n))$ means that there exists a positive constant $c_{2}$
and an integer $m_{2}$ such that $f(n)\geq c_{2}g(n)$ for all $n\geq m_{2}$\emph{.
}Namely,\emph{ $g(n)=O(f(n))$.}
\item $f(n)=\Theta(g(n))$ means that both $f(n)=O(g(n))$ and $f(n)=\Omega(g(n))$
hold for all $n\geq\textrm{max}\left(m_{1},\, m_{2}\right)$.
\end{enumerate}

\subsection{Network Model}

Consider the scenario where a network of primary nodes and a network
of secondary nodes coexist over a unit square. The primary nodes are
distributed according to a Poisson point process (P.~P.~P.) of density
$n$ and randomly grouped into one-to-one source-destination (S-D)
pairs. The distribution of the secondary nodes is following a P.~P.~P.
of density $m$. The secondary nodes are also randomly grouped into
one-to-one S-D pairs. As the model in~\cite{Scaling:Jeon}, we assume
that the density of the secondary network is higher than that of the
primary network, i.e.,\begin{equation}
m=n^{\beta},\label{Eq1}\end{equation}
with $\beta>1$.

For the wireless channel, we only consider the large-scale pathloss
and ignore the effects of shadowing and small-scale multipath fading.
As such, the normalized channel power gain $g(r)$ is given as\begin{equation}
g(r)=\frac{A}{r^{\alpha}},\label{Eq2}\end{equation}
where $A$ is a system-dependent constant, $r$ is the distance between
the TX and the corresponding RX, and $\alpha>2$ denotes the pathloss
exponent. In the following discussion, we normalize $A$ to be unity
for simplicity. 

The primary network and the secondary network share the same spectrum,
time, and space, while the former one is the licensed user of the
spectrum and thus has a higher priority to access the spectrum. The
secondary network opportunistically access the spectrum while keeping
its interference to the primary network at an {}``acceptable level''.
In this paper, the {}``acceptable level'' means that the presence
of the secondary network does not degrade the throughput scaling law
of the primary network.

We assume that the secondary network only knows the locations of the
primary TXs and has no knowledge about the locations of the primary
RXs. This is the essential difference between our model and the model
in~\cite{Scaling:Jeon}, where the authors assumed that the secondary
network knows the locations of all the primary nodes. Some other aspects
of our model are defined in a similar way to that in~\cite{Scaling:Jeon},
as we will discuss later.

\subsection{Transmission Rate and Throughput}

The ambient noise is assumed as additive white Gaussian noise (AWGN)
with an average power $N_{0}$. During each transmission, we assume
that each TX-RX pair deploys a capacity-achieving scheme, and the
channel bandwidth is normalized to be unity for simplicity. Thus the
data rate of the $k$-th primary TX-RX pair is given by \begin{equation}
R_{p}(k)=\log\left(1+\frac{P_{p}(k)g\left(\Vert X_{p,\textrm{tx}}(k)-X_{p,\textrm{rx}}(k)\Vert\right)}{N_{0}+I_{p}(k)+I_{sp}(k)}\right),\label{Eq3}\end{equation}
where $\Vert\cdot\Vert$ stands for the norm operation, $P_{p}(k)$
is the transmit power of the $k$-th primary TX-RX pair, $X_{p,\textrm{tx}}(k)$
and $X_{p,\textrm{rx}}(k)$ are the TX and RX locations of the $k$-th
primary TX-RX pair, respectively, $I_{p}(k)$ is the sum interference
from all other primary TXs to the RX of the $k$-th primary TX-RX
pair, and $I_{sp}(k)$ is the sum interference from all the secondary
TXs to the RX of the $k$-th primary TX-RX pair. Specifically, $I_{p}(k)$
can be written as

\begin{equation}
I_{p}(k)={\displaystyle \sum_{i=1,i\neq k}^{Q_{p}}P_{p}(k)g\left(\parallel X_{p,\textrm{tx}}(i)-X_{p,\textrm{rx}}(k)\parallel\right),}\label{eq:Ip}\end{equation}
where $Q_{p}$ is the number of active primary TX-RX pairs, and $I_{sp}(k)$
is given by

\begin{equation}
I_{sp}(k)={\displaystyle \sum_{i=1}^{Q_{s}}P_{s}(i)g\left(\parallel X_{s,\textrm{tx}}(i)-X_{p,\textrm{rx}}(k)\parallel\right),}\label{eq:Isp}\end{equation}
where $Q_{s}$ is the number of active secondary TX-RX pairs, $P_{s}(i)$
is the transmit power of the $i$-th secondary TX-RX pair, and $X_{s,\textrm{tx}}(i)$
is the TX location of the $i$-th secondary TX-RX pair. Likewise,
the data rate of the $l$-th secondary TX-RX pair is given by

\begin{equation}
R_{s}(l)=\log\left(1+\frac{P_{s}(l)g\left(\Vert X_{s,\textrm{tx}}(l)-X_{s,\textrm{rx}}(l)\Vert\right)}{N_{0}+I_{s}(l)+I_{ps}(l)}\right),\label{Eq4}\end{equation}
where $X_{s,\textrm{rx}}(l)$ is the RX location of the $l$-th secondary
TX-RX pair, $I_{s}(l)$ is the sum interference from all other secondary
TXs to the RX of the $l$-th secondary TX-RX pair, and $I_{ps}(l)$
is the sum interference from all primary TXs to the RX of the $l$-th
secondary TX-RX pair. Specifically, $I_{s}(l)$ is given by

\begin{equation}
I_{s}(l)={\displaystyle \sum_{i=1,i\neq l}^{Q_{s}}P_{s}(i)g\left(\parallel X_{s,\textrm{tx}}(i)-X_{s,\textrm{rx}}(l)\parallel\right),}\label{eq:Is}\end{equation}
and $I_{ps}(l)$ is given by\begin{equation}
I_{ps}(l)={\displaystyle \sum_{i=1}^{Q_{p}}P_{p}(i)g\left(\parallel X_{p,\textrm{tx}}(i)-X_{s,\textrm{rx}}(l)\parallel\right).}\label{eq:Ips}\end{equation}

Now we give the definitions of throughput per S-D pair and sum throughput.

\begin{definitn}
\label{Df:Th_pernode}The \emph{throughput per S-D pair} $\lambda(n_{t})$
is defined as the average data rate that each source node can transmit
to its chosen destination $w.h.p.$ in a multi-hop fashion with a
particular scheduling scheme, where $n_{t}$ is the number of nodes
in the network. We have

\begin{equation}
p\left(\underset{1\leq i\leq n_{t}/2}{\textrm{min}}\underset{t\rightarrow\infty}{\liminf}\frac{1}{t}M_{i}(t)\geq\lambda(n_{t})\right)\rightarrow1,\label{eq:Throughput}\end{equation}
as $n_{t}\rightarrow\infty$, where $M_{i}(t)$ is the number of bits
that S-D pair $i$ transmitted in $t$ time slots. 
\begin{definitn}
\label{Df:Th_sum}The \emph{sum throughput} $T(n_{t})$ is defined
as the product between the throughput per S-D pair $\lambda(n_{t})$
and the number of S-D pairs in the network, i.e.,

\begin{equation}
T(n_{t})=\frac{n_{t}}{2}\lambda(n_{t}).\end{equation}

\end{definitn}
\end{definitn}
~~~~According to the network model defined in Section II.A, the
number of nodes in the primary network (or in the secondary network)
is a random variable. However, we will show in Lemma~\ref{Lem2}
and Lemma~\ref{lem:SNumber} at Section III that the number of nodes
in the primary network (or in the secondary network) will be bounded
by functions of the node density \emph{w.h.p.}. As such, in the following
discussion, we use $\lambda_{p}(n)$ and $\lambda_{s}(m)$ to denote
the throughputs per S-D pair for the primary network and the secondary
network, respectively. We use $T_{p}(n)$ and $T_{s}(m)$ to denote
the sum throughputs for the primary network and the secondary network,
respectively.

\subsection{Fluid Model and Delay}

As in~\cite{Delay:Gamal1}, we use a fluid model to study the delay-throughput
tradeoffs for the primary and secondary networks. In this model, we
divide each time slot into multiple packet slots, and the size of
the data packets can be scaled down to arbitrarily small with the
increase of the node density $n$ (or $m$) in the networks. 

\begin{definitn}
The \emph{delay} $D(n_{t})$ \emph{of a packet} is defined as the
average time that it takes to reach the destination node after the
departure from the source node.
\end{definitn}
~~~~Let $D_{i}(j)$ denote the delay of packet $j$ for S-D pair
$i$. The sample mean of delay over all packets transmitted for S-D
pair $i$ is defined as

\begin{equation}
D_{i}=\underset{k\rightarrow\infty}{\limsup}\frac{1}{k}\sum_{j=1}^{k}D_{i}(j),\label{eq:delay1}\end{equation}
and the average delay over all S-D pairs is given by\[
\overline{D(n_{t})}=\frac{2}{n_{t}}\sum_{i=1}^{n_{t}/2}D_{i}.\]
The average delay over all realizations of the network is\begin{equation}
D(n_{t})=E\left[\,\overline{D(n_{t})}\,\right]=\frac{2}{n_{t}}\sum_{i=1}^{n_{t}/2}E\left[D_{i}\right].\label{eq:delay2}\end{equation}

~~As what we did over the notations of throughput, in the following
discussion, we use $D_{p}(n)$ and $D_{s}(m)$ to denote the packet
delays for the primary network and the secondary network, respectively.

\subsection{Main Results}

The main results of this paper are as follows.

\begin{enumerate}
\item We propose a coexistence scheme for two overlaid ad hoc wireless networks:
a primary network vs. a secondary network. These two networks operate
in the same geographic region and share the same spectrum. The primary
network has a higher priority to access the spectrum and has no special
considerations over the presence of the secondary network, while the
secondary network operates opportunistically to access the spectrum
in order to limit the interference to the primary network. We assume
that the primary network uses a typical time-slotted adjacent-neighbor
transmission protocol (similar to that in~\cite{Capacity:Gupta})
and the secondary network has a higher density and only knows the
locations of the primary TXs. By a properly designed secondary protocol,
we show that each secondary source node has a finite opportunity to
transmit its packets to the chosen destination\emph{ w.h.p.}, i.e.,
no outage compared with the result in~\cite{Scaling:Jeon}. 
\item For the primary network, we show that the throughput per S-D pair
is $\lambda_{p}(n)=\Theta(\sqrt{\frac{1}{n\log n}})$ \emph{w.h.p}.
and the sum throughput is $T_{p}(n)=\Theta(\sqrt{\frac{n}{\log n}})$
\emph{w.h.p}.. These results are the same as those in a stand-alone
ad hoc wireless network considered in~\cite{Capacity:Gupta}. Following
the fluid model~\cite{Delay:Gamal1}, we give the delay-throughput
tradeoff for the primary network as $D_{p}(n)=\Theta(n\lambda_{p}(n))$
for $\lambda_{p}(n)=O(\frac{1}{\sqrt{n\log n}})$, which is the optimal
delay-throughput tradeoff for a stand-alone wireless ad hoc network
established in~\cite{Delay:Gamal1}. 
\item For the secondary network, we prove that the throughput per S-D pair
is $\lambda_{s}(m)=\Theta(\sqrt{\frac{1}{m\log m}})$ \emph{w.h.p}.
and the sum throughput is $T_{s}(m)=\Theta(\sqrt{\frac{m}{\log m}})$
\emph{w.h.p}.. Although due to the presence of the preservation regions,
the secondary packets seemingly experience larger delays compared
with that of the primary network, we show that the delay-throughput
tradeoff for the secondary network is the same as that in the primary
network, i.e., $D_{s}(m)=\Theta(m\lambda_{s}(m))$ for $\lambda_{s}(m)=O(\frac{1}{\sqrt{m\log m}})$.
\end{enumerate}

\section{Network Protocols}

In our proposed scheme, the primary network deploys a modified time-slotted
multi-hop transmission scheme over that in~~\cite{Capacity:Gupta},
\cite{Delay:Gamal1}, \cite{Scaling:Jeon}. The secondary network
adapts its protocol according to the primary transmission scheme.
We first describe the primary protocol, then introduce the secondary
protocol, and finally give a lemma to show that with our proposed
protocols the secondary users can communicate without outage \emph{w.h.p.}.
Similarly as in~\cite{Scaling:Jeon}, we claim that an outage event
occurs when a node has zero opportunity to communicate. The outage
probability is defined as the fraction of nodes that have zero opportunity
to communicate.

\begin{figure}
\begin{centering}
\includegraphics[width=2.5in]{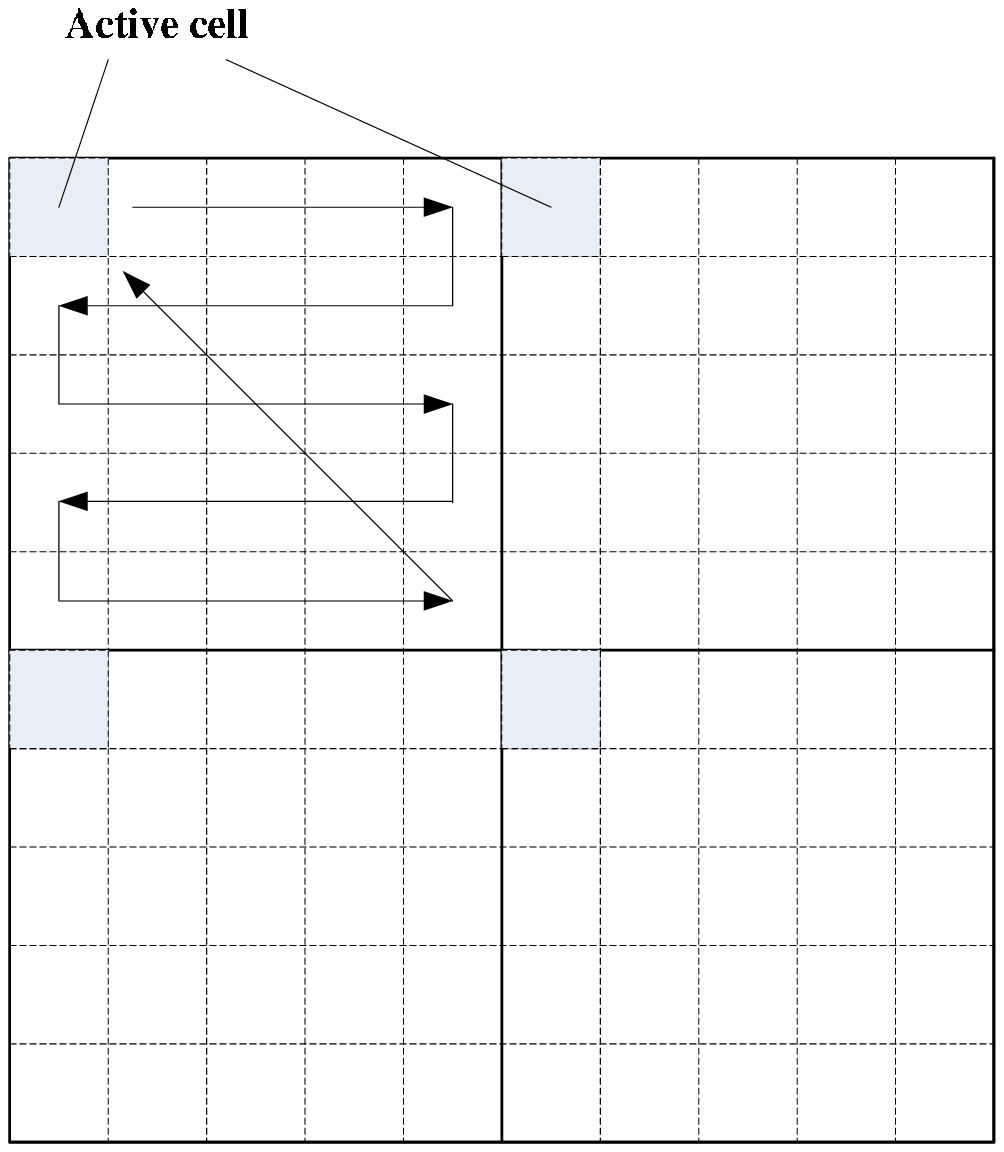}
\par\end{centering}

\caption{\label{Fig11}A four-cluster example with 25 cells per cluster. The
cells in each cluster take turns to be active along the arrowed line
over time.}

\end{figure}

\subsection{Primary Network Protocol}

\begin{figure}
\begin{centering}
\centering \includegraphics[width=2.5in]{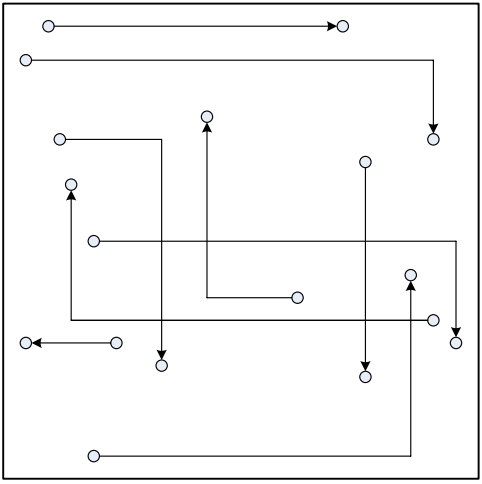} 
\par\end{centering}

\caption{\label{Fig1} Examples of HDPs and VDPs for the primary S-D pairs.}

\end{figure}

\begin{figure}
\begin{centering}
\includegraphics[angle=-90,width=2.5in]{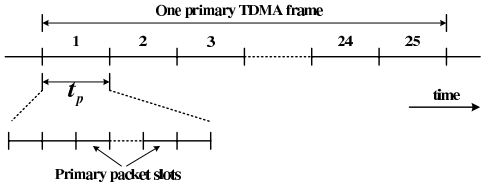}
\par\end{centering}

\caption{\label{Fig17}Structure of the primary TDMA frame, where $t_{p}$
is the time-slot duration of the primary TDMA scheme.}

\end{figure}

\begin{itemize}
\item We divide the unit square into small-square primary cells. The area
of each primary cell is $a_{p}=\frac{k_{1}\log n}{n}$, with $k_{1}\geq1$. 
\item We group the primary cells into primary clusters, and each cluster
has $K_{p}^{2}=25$ primary cells. We split the transmission time
into time division multiple access (TDMA) frames, where each frame
has 25 time slots that correspond to the number of cells in each primary
cluster with each slot of length $t_{p}$. In each time slot, one
cell in each primary cluster is chosen to be active. The cells in
each primary cluster take turns to be active in a round-robin fashion.
All primary clusters follow the same 25-TDMA transmission pattern,
as shown in Fig.~\ref{Fig11}. 
\item We define the data path along which the packets traverse as the horizontal
line and then the vertical line connecting a source and its corresponding
destination, as shown in Fig.~\ref{Fig1}. One node within a primary
cell is defined as a designated relay node, which is responsible for
relaying the packets of all the data paths passing through the cell.
The packets will be forwarded from cell to cell by the relay nodes
first along the horizontal data path (HDP), then along the vertical
data path (VDP). Nodes in a particular cell take turns to serve as
the designated relay node. 
\item When a primary cell is active, it transmits a single packet for each
of the data paths passing through the cell. The transmission is also
deployed in a TDMA fashion. The TDMA frame structure for the primary
network is shown in Fig.~\ref{Fig17}, where one packet slot is assigned
to one S-D data path that passes through or originates from a particular
primary cell. As such, the number of packet slots is determined by
the total number of data paths in the cell, which is based on the
so-called fluid model~\cite{Delay:Gamal1}. The specific packet transmission
procedure is as follows:

\begin{itemize}
\item The designated relay node first transmits a single packet for each
of the S-D paths passing through the cell; and then each of the source
nodes within the cell takes turns to transmit a single packet. 
\item The receiving node must be located in one of the neighboring primary
cells along the predefined data path, unless it is a destination node,
which may be located in the same cell. If the next-hop of the packet
is the final destination, it will be directly delivered to the destination
node; otherwise, the packet will be transmitted to a designated relay
node. 
\item The designated relay node in each primary cell maintains a buffer
to temporarily store the packets received from its neighboring cells,
and each packet will be transmitted to the next hop in the next active
time slot of the cell. 
\end{itemize}
\item At each packet slot, the TX node transmits with power of $P_{0}a_{p}^{\frac{\alpha}{2}}$,
where $P_{0}$ is a constant. 
\end{itemize}

The primary protocol in this paper is similar to that in~\cite{Delay:Gamal1}
but with different data paths and TDMA transmission patterns. As a
result, we have the following two lemmas. 

\begin{lemma}
\label{Lem2} Let $n_{pt}$ denote the number of total primary nodes
in the unit square; then we have $\frac{n}{2}<n_{pt}<en$\emph{ w.h.p.}.
\end{lemma}
\begin{proof}
Since $n_{pt}$ is a Poisson random variable with parameter $\mu=n$,
using the Chernoff bound in Lemma~\ref{lem:Mitz} (see Appendix),
we have\begin{eqnarray}
p\left(n_{pt}\leq\frac{n}{2}\right) & \leq & \frac{e^{-n}(en)^{\frac{n}{2}}}{\left(\frac{n}{2}\right)^{\frac{n}{2}}}\nonumber \\
 & = & \left(\frac{2}{e}\right)^{\frac{n}{2}}\rightarrow0\label{eq:primarynode}\end{eqnarray}
as $n\rightarrow\infty$, and\begin{eqnarray}
p\left(n_{pt}\geq en\right) & \leq & \frac{e^{-n}(en)^{en}}{(en)^{en}}\nonumber \\
 & = & e^{-n}\rightarrow0\label{eq:primarynode1}\end{eqnarray}
as $n\rightarrow\infty$. Combining (\ref{eq:primarynode}) and (\ref{eq:primarynode1})
via the union bound, we obtain\[
p\left(n_{pt}\leq\frac{n}{2}\:\textrm{or}\: n_{pt}\geq en\right)\leq p\left(n_{pt}\leq\frac{n}{2}\right)+p\left(n_{pt}\geq en\right)\rightarrow0\]
as $n\rightarrow\infty$. Hence\[
p\left(\frac{n}{2}<n_{pt}<en\right)=1-p\left(n_{pt}\leq\frac{n}{2}\:\textrm{or}\: n_{pt}\geq en\right)\rightarrow1\]
as $n\rightarrow\infty$, which completes the proof.
\end{proof}
\begin{lemma}
\label{lem:Number}For $k_{1}\geq1,$ each primary cell contains at
least one but no more than $k_{1}e\log n$ primary nodes \emph{w.h.p.}.
\end{lemma}
\begin{proof}
Let $n_{p}$ denote the number of primary nodes in a particular primary
cell; then $n_{p}$ is a Poisson random variable with parameter $\mu=na_{p}=k_{1}\log n$.
The probability of $n_{p}=0$ is given by

\begin{equation}
p\left(n_{p}=0\right)=\frac{e^{-k_{1}\log n}(k_{1}\log n)^{k}}{k!}\biggl|_{k=0}=\frac{1}{n^{k_{1}}}.\end{equation}
By the union bound, the probability that at least one primary cell
having no nodes is upper-bounded by the total number of cells multiplied
by $p\left(n_{p}=0\right)$, which is\begin{align*}
p\left(\textrm{At least one primary cell has no nodes}\right)\\
\leq\frac{1}{a_{p}}p\left(n_{p}=0\right)=\frac{1}{k_{1}n^{k_{1}-1}\log n}\rightarrow0\quad\:\end{align*}
as $n\rightarrow\infty$ for $k_{1}\geq1$.

Now consider the upper bound of $n_{p}$. By the Chernoff bound in
Lemma~\ref{lem:Mitz} (see Appendix), we have\[
p\left(n_{p}\geq k_{1}e\log n\right)\leq\frac{e^{-k_{1}\log n}\left(ek_{1}\log n\right)^{k_{1}e\log n}}{\left(k_{1}e\log n\right)^{k_{1}e\log n}}=n^{-k_{1}}.\]
As long as $k_{1}\geq1$, by the union bound, we have\begin{align*}
p\left(\textrm{At least one primary cell has more than }k_{1}e\log n\,\textrm{nodes}\right)\\
\leq\frac{1}{a_{p}}n^{-k_{1}}=\frac{1}{k_{1}n^{k_{1}-1}\log n}\rightarrow0\qquad\qquad\qquad\qquad\end{align*}
as $n\rightarrow\infty$. This completes the proof.
\end{proof}

\subsection{Secondary Network Protocol}

\begin{figure}
\begin{centering}
\includegraphics[angle=-90,width=2.7in]{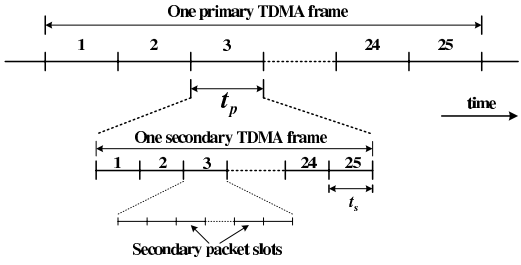} 
\par\end{centering}

\caption{\label{Fig6}Structure of the secondary TDMA frame and its relationship
with the primary TDMA frame, where $t_{s}$ is the time-slot duration
for the secondary TDMA scheme.}

\end{figure}

\begin{figure}
\begin{centering}
\includegraphics[width=2.8in]{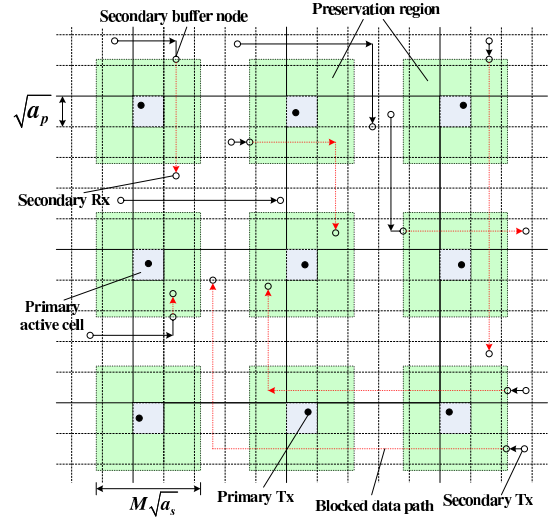} 
\par\end{centering}

\caption{\label{Fig2}Preservation region and examples of secondary data paths.}

\end{figure}

\begin{itemize}
\item We divide the unit area into square secondary cells with size $a_{s}=\frac{k_{2}\log m}{m}$,
with $k_{2}$$\geq1$. 
\item We group the secondary cells into secondary clusters. Each secondary
cluster has $K_{s}^{2}=25$ cells. Similar to the primary network
protocol, the secondary network also follows a 25-TDMA pattern to
communicate. We let the duration of each secondary TDMA frame equal
to that of one primary time slot. The relationship between the primary
TDMA frame and the secondary TDMA frame is shown in Fig.~\ref{Fig6},
where each secondary time slot is further divided into packet slots.
\item To limit the interference from the secondary nodes to the primary
nodes, we define a preservation region as a square containing $M^{2}$
secondary cells around a particular primary cell in which an active
primary TX (not the RX) is located, where $M$ is an integer and the
value will be defined later. No secondary nodes in the preservation
regions are allowed to transmit.
\item The designated relay nodes and data paths for the secondary network
are defined in the same way as those for the primary network. As shown
in Fig.~\ref{Fig2}, when a particular secondary cell outside the
preservation region is active, its designated relay node transmits
a single packet for each of the data paths passing through the cell,
and each of the secondary source nodes within the cell takes turns
to transmit a single packet. The packet is transmitted to the next-hop
relay node or the destination node in neighboring secondary cells
along the HDP or VDP path. Note that if the RX node is the destination
node, it may be located in the same cell, as we discussed for the
primary protocol.
\item When a secondary cell falls into a preservation region%
\footnote{Note that the secondary nodes located in the preservation regions
can still receive packets from TXs outside the preservation regions,
although they are not permitted to transmit packets.%
}, its designated relay node buffers the packets that it receives;
it waits until the preservation region is cleared and the cell is
active to deliver the packets to the next hop. 
\item At each packet slot, the active secondary TX node transmits with power
of $P_{1}a_{s}^{\frac{\alpha}{2}}$, where $P_{1}$ is a constant.
\end{itemize}
~~~~Similarly as in the primary network case, we have the following
two lemmas for the secondary network.

\begin{lemma}
\label{lem:SNumber}Let $n_{st}$ denote the total number of secondary
nodes in the unit square; then we have $\frac{m}{2}<n_{st}<em$\emph{
w.h.p.}.
\end{lemma}
\begin{proof}
The proof is similar to that of Lemma~\ref{Lem2}.
\end{proof}
\begin{lemma}
For $k_{2}\geq1,$ each secondary cell contains at least one but no
more than $k_{2}e\log m$ secondary nodes \emph{w.h.p.}.
\end{lemma}
\begin{proof}
The proof is similar to that of Lemma~\ref{lem:Number}.
\end{proof}
~~Regarding the cluster size, note that the value of $K_{s}$ is
not necessarily the same as that of $K_{p}$. Here we choose $K_{s}=K_{p}$
for simplicity. Without loss of generality, we also choose $k_{1}=k_{2}$
in the following discussion. 

Now, let us discuss how to choose the value of $M$, i.e., the size
for the preservation region. Considering the fact that the primary
TX may only transmit to a node in its adjacent cells or within the
same cell, the preservation region should accommodate at least 9 primary
cells to protect the potential primary RX. Since the primary RX may
be located close to the outer boundary of the 9-cell region, we should
add another layer of protective secondary cells. As such, any active
secondary TXs outside the preservation region are at least certain-distance-away
from the potential primary RX. Therefore, we define the side length
of the preservation square region as

\begin{equation}
M\sqrt{a_{s}}\geq3\sqrt{a_{p}}+2\epsilon_{p},\label{eq:Preservation}\end{equation}
where $\epsilon_{p}>0$ defines the width of the protective secondary
strip around the 9 primary cells in the preservation region. There
is a tradeoff in choosing the value of $\epsilon_{p}$. If we choose
a larger $\epsilon_{p}$, the interference from the secondary network
to the primary network will be less. However, the opportunity for
the secondary network to access the spectrum will also be less since
the unpreserved area in the unit square will be reduced. In the following
discussion, we set $\epsilon_{p}=\sqrt{a_{s}}$ for simplicity. Accordingly,
the minimum value of $M$ can be set as\begin{eqnarray}
M & = & \lfloor\frac{3\sqrt{a_{p}}+2\sqrt{a_{s}}}{\sqrt{a_{s}}}\rfloor\nonumber \\
 & = & \lfloor3\sqrt{\frac{a_{p}}{a_{s}}}\rfloor+2\nonumber \\
 & \approx & 3\sqrt{\frac{n^{\beta-1}}{\beta},}\label{Eq5a}\end{eqnarray}

\begin{flushleft}
where $\lfloor\cdot\rfloor$ denotes the flooring operation. In the
last equation of~(\ref{Eq5a}), we applied $a_{p}=\frac{k_{1}\log n}{n}$,
$a_{s}=\frac{k_{2}\log m}{m}$, $k_{1}=k_{2}$, and~(\ref{Eq1}),
assuming that $n$ is large enough. In the following discussion, {}``$n$
is large'' or {}``$n$ is large enough'' means that, for a fixed
$\beta$, $n$ is chosen to satisfy $a_{s}\ll a_{p}$. For example,
when $k_{1}=k_{2}$, $\beta=2$, $n=1000$, we have $m=1000000$ and
$\frac{a_{p}}{a_{s}}=\frac{n^{\beta-1}}{\beta}=500$.
\par\end{flushleft}

~~Note that the preservation region defined here is larger than
that in~\cite{Scaling:Jeon} due to the fact that we only know the
locations of primary TXs. If a secondary node falls inside a preservation
region, it will be silenced. If not, it may become active and has
an opportunity to transmit its packets. Accordingly, we call the unpreserved
region as the {}``active region''. Since the locations of preservation
regions change periodically according to the active time slots in
the primary TDMA frame, from the point view of a specific secondary
node, it is periodically located in the active region. We define the
following terminology to measure the fraction of time in which a secondary
cell is located in the active region.

\begin{definitn}
\label{Df:Active} The \emph{opportunistic factor} of a secondary
cell is defined as the fraction of time in which it is located in
the active region.
\end{definitn}
~~We use the following lemma to show that, with the protocols defined
previously, each individual secondary source node has a finite opportunity
to transmit its packets to the chosen destination \emph{w.h.p.}.

\begin{figure}
\begin{centering}
\includegraphics[angle=-90,width=2.5in]{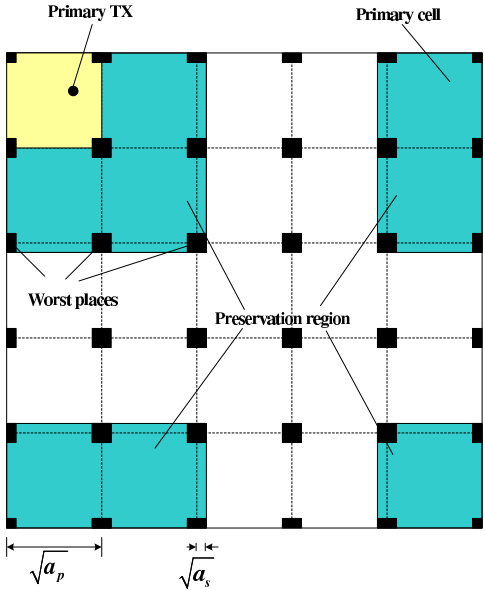} 
\par\end{centering}

\begin{centering}
\caption{\label{Fig:Cluster}Preservation regions and worst places in one primary
cluster.}

\par\end{centering}
\end{figure}

\begin{figure}
\begin{centering}
\includegraphics[angle=-90,width=2.5in]{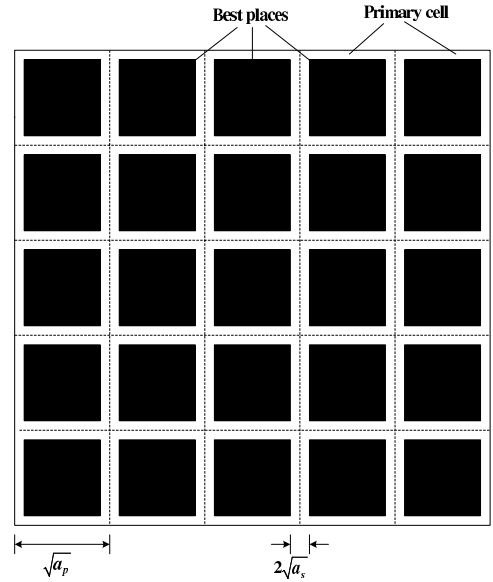}
\par\end{centering}

\caption{\label{Fig:Best}The best places in one primary cluster.}

\end{figure}

\begin{lemma}
\label{Lem:Active} With the proposed transmission protocol, we have
the following results: 
\begin{enumerate}
\item The opportunistic factor for a secondary cell is $\frac{9}{25}\leq\eta\leq\frac{16}{25}$,
for $n$ is large enough. 
\item Each individual secondary node has a finite opportunity to transmit
its packets to the chosen destination, i.e., zero outage, \emph{w.h.p.}. 
\end{enumerate}
\end{lemma}
\begin{proof}
Consider one primary cluster of 25 primary cells as shown in Fig.~\ref{Fig:Cluster},
where the preservation regions are illustrated as the shaded area
when the upper-left primary cell is active in this and neighboring
clusters. The primary cells will take turns to be active over time
(see Fig.~\ref{Fig11}) and the locations of the preservation regions
will change accordingly. We can easily verify that any point in the
cluster has a finite opportunity to be in the active region when $n$
is large. However, during each period of a primary TDMA frame, the
fractions of time for different secondary nodes to be in the active
region are not the same. The worst places are the squares with side
length of $2\sqrt{a_{s}}$ around the vertices of each primary cell,
as shown by those deeply-shaded small squares in Fig.~\ref{Fig:Cluster}.
The opportunistic factor of the secondary cells in these squares is
$\frac{9}{25}$. The best places are the squares with side length
of $\sqrt{a_{p}}-2\sqrt{a_{s}}$ inside each primary cell, as shown
by the deeply-shaded squares in Fig.~\ref{Fig:Best}. The opportunistic
factor of the secondary cells in these squares are $\frac{16}{25}$.
When the secondary cell lies in other places, the opportunistic factor
is between $\frac{9}{25}$ and $\frac{16}{25}$. 

The condition that a secondary node is located in the active region
is not sufficient to ensure that it can transmit packets to the destination
along the predefined data path. Recall that the secondary network
also deploys a TDMA scheme with adjacent-neighbor transmission. The
sufficient condition to ensure that each individual secondary node
has a finite chance to transmit packets is that the secondary cell
in which the node is located will be assigned with at least one active
secondary TDMA slot within each secondary frame, whenever the cell
is in the active region. Since in each primary time slot, we have
one complete secondary TDMA frame in our protocol, the above sufficient
condition is indeed satisfied. 

Based on the above discussions, during each period of a primary TDMA
frame, each secondary cell has a finite opportunity to be located
in the active region with an opportunistic factor of $\frac{9}{25}\leq\eta\leq\frac{16}{25}$,
and each of them is assigned with a secondary TDMA slot. According
to the secondary protocol, when a secondary cell is active, each packet
buffered in this cell will be assigned with a packet slot\emph{ w.h.p}.
to be transmitted, since the total number of data paths that pass
through or originate from each secondary cell is upper-bounded \emph{w.h.p.}
(see Lemma~\ref{lem:SecondaryDP} in Section~V). Thus, the packets
from any secondary source node have a finite opportunity to be transmitted
along the predefined data path to the chosen destination\emph{~w.h.p.}.
This completes the proof for the zero outage property. 
\end{proof}
~~There is a significant difference between our result here and
that in~\cite{Scaling:Jeon}. The authors in~\cite{Scaling:Jeon}
defined preservation regions of 9 secondary cells around each primary
node, and the positions of the preservation regions are fixed. If
the secondary nodes are located in the preservation regions, they
will never be active. Therefore, the secondary network in~\cite{Scaling:Jeon}
usually suffers from a non-zero outage probability, even though the
outage probability is upper-bounded \emph{w.h.p.}. In our case, each
secondary node has a finite opportunity to be active such that we
have zero outage \emph{w.h.p.}.

\section{Delay and Throughput Analysis for the Primary Network }

In this section, we discuss the delay and throughput scaling laws
as well as the delay-throughput tradeoff for the primary network.
The main results are given in three theorems. We first present the
delay and throughput scaling laws, then establish the delay-throughput
tradeoff for the primary network.

\subsection{Delay Analysis for the Primary Network}

The packet delay for the primary network is given by the following
theorem.

\begin{thm}
\label{thm:PDelay}According to the primary network protocol in Section
III, the packet delay is given by

\begin{equation}
D_{p}(n)=\Theta\left(\frac{1}{\sqrt{a_{p}(n)}}\right),\quad w.h.p..\label{eq:P_Delay}\end{equation}

\end{thm}
\begin{proof}
We first derive the average number of hops for each packet to traverse
along the primary S-D data path, then use the fact that the time for
each primary packet to spend at each hop is a constant, $25t_{p}$,
as shown in Fig.~\ref{Fig17}, and finally calculate the average
delay for each primary S-D pair . 

Since each primary hop spans a distance of $\Theta\left(\sqrt{a_{p}(n)}\right)$\emph{
w.h.p.}, the number of hops for a primary packet along the S-D data
path $i$ is $\Theta\left(\tfrac{d_{p}(i)}{\sqrt{a_{p}(i)}}\right)$
\emph{w.h.p.}, where $d_{p}(i)$ is the length of the primary S-D
data path $i$. Hence, the number of hops traversed by a primary packet,
averaged over all S-D pairs, is $\Theta\left(\frac{2}{n_{pt}}{\textstyle \sum_{i=1}^{n_{pt}/2}\tfrac{d_{p}(i)}{\sqrt{a_{p}(n)}}}\right)$\emph{
w.h.p}..

The data path length $d_{p}(i)$ is a random variable, with a maximum
value of 2. According to the law of large numbers, as $n_{pt}\rightarrow\infty$,
the average distance between primary S-D pairs is 

\[
\frac{2}{n_{pt}}{\displaystyle \sum_{i=1}^{n_{pt}/2}d_{p}(i)=\Theta\left(1\right)}.\]

Therefore, the average number of hops for a primary packet to traverse
is $\Theta\left(\tfrac{1}{\sqrt{a_{p}(n)}}\right)$ \emph{w.h.p.}.
Since we use a fluid model such that the packet size of the primary
network scales proportionally to the throughput $\lambda_{p}(n)$,
each packet arrived at a primary cell will be transmitted in the next
active time slot of the cell. As such, the maximum time spent at each
primary hop for a particular packet is $25t_{p}$. Hence, the average
delay for each primary packet is given by 

\begin{equation}
D_{p}(n)=\Theta\left(\frac{25t_{p}}{\sqrt{a_{p}(n)}}\right)=\Theta\left(\frac{1}{\sqrt{a_{p}(n)}}\right),\quad w.h.p,\label{eq:DelayPR}\end{equation}
 which completes the proof.
\end{proof}
~~The above proof follows the same logic as the proof of Theorem
4 in~\cite{Delay:Gamal1}. The two differences are that we use HDPs
and VDPs as the packet routing paths instead of the direct S-D links
and we use a different TDMA transmission pattern.

\subsection{Throughput Analysis for the Primary Network}

For the primary network, the throughput per S-D pair and the sum throughput
scaling laws are given in the following theorem.

\begin{thm}
\label{Th1}With the primary protocol defined in Section III, the
primary network can achieve the following throughput per S-D pair
and sum throughput \emph{w.h.p.}: \begin{equation}
\lambda_{p}(n)=\Theta\left(\sqrt{\frac{1}{n\log n}}\right)\label{Eq9}\end{equation}
and \begin{equation}
T_{p}(n)=\Theta\left(\sqrt{\frac{n}{\log n}}\right).\label{Eq10}\end{equation}

\end{thm}
~~Before we give the proof of the above theorem, we first give two
lemmas, then use these lemmas to prove the theorem. The main logical
flows in the proofs of these lemmas and the theorem are motivated
by that in~\cite{Scaling:Jeon} and~\cite{Delay:Toumpis}. 

\begin{figure}
\begin{centering}
\includegraphics[width=2.8in]{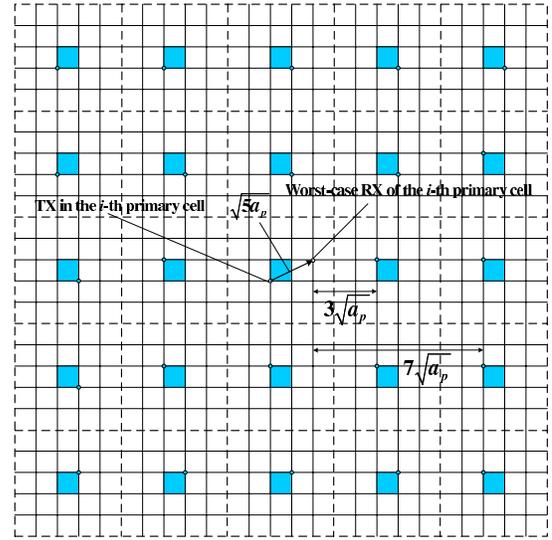} 
\par\end{centering}

\begin{centering}
\caption{\label{Fig5} Interference from the concurrent primary transmissions
to the worst-case primary RX of the transmission from the $i$-th
primary cell.}

\par\end{centering}
\end{figure}

\begin{lemma}
\label{Lem4} With the primary protocol defined in Section III, each
TX node in a primary cell can support a constant data rate of $K_{1}$,
where $K_{1}>0$ is independent of $n$. 
\end{lemma}
\begin{proof}
In a given primary packet slot, suppose we have $Q_{p}$ active primary
cells and $Q_{s}$ active secondary cells. The data rate supported
for a TX node in the $i$-th active primary cell can be calculated
as follows: \begin{equation}
R_{p}(i)=\frac{1}{25}\log\left(1+\frac{P_{p}(i)g(\|X_{p,\textrm{tx}}(i)-X_{p,\textrm{rx}}(i)\|)}{N_{0}+I_{p}(i)+I_{sp}(i)}\right),\end{equation}
where $\frac{1}{25}$ denotes the rate loss due to the 25-TDMA transmission
in the primary network. Note that since there is only one active primary
link initiated in each primary cell at a given time, we index the
active link initiated in the $i$-th active primary cell as the $i$-th
active primary link in the whole network. In Fig.~\ref{Fig5}, we
show the primary interference sources to the primary RX of the $i$-th
active primary link, where the shaded cells represent the active primary
cells based on the 25-TDMA protocol. From the figure, we see that
we have 8 primary interferers with a distance of at least $3\sqrt{a_{p}}$,
16 primary interferers with a distance of at least $7\sqrt{a_{p}}$,
and so on. Thus, $I_{p}(i)$ is upper-bounded as\begin{eqnarray}
I_{p}(i) & = & \sum_{k=1,k\neq i}^{Q_{p}}P_{p}(k)g(\|X_{p,\textrm{tx}}(k)-X_{p,\textrm{rx}}(i)\|)\nonumber \\
 & < & P_{0}\sum_{t=1}^{\infty}8t(4t-1)^{-\alpha}\nonumber \\
 & = & I_{p}<\infty,\label{eq:Ip1}\end{eqnarray}
where we used the relationship that $P_{p}(k)=P_{0}a_{p}^{\frac{\alpha}{2}}$
for all $k$'s and the fact that the series $\sum_{t=1}^{\infty}8t(4t-1)^{-\alpha}$
converges to a constant for $\alpha>2$ (see Lemma~\ref{Lem5} in
Appendix). Due to the preservation regions, a minimum distance $\sqrt{a_{s}}$
can be guaranteed from all secondary active TXs to any active primary
RXs. Thus, $I_{sp}(i)$ is upper-bounded as\begin{eqnarray}
I_{sp}(i) & = & \sum_{k=1}^{Q_{s}}P_{s}(k)g(\|X_{s,\textrm{tx}}(k)-X_{p,\textrm{rx}}(k)\|)\nonumber \\
 &  & +P_{1}a_{s}^{\frac{\alpha}{2}}\left(\sqrt{a_{s}}\right)^{-\alpha}\nonumber \\
 & < & P_{1}\sum_{t=1}^{\infty}8t(4t-1)^{-\alpha}+P_{1}\nonumber \\
 & = & I_{sp}<\infty,\label{eq:Eq24}\end{eqnarray}
 where we used the fact that $P_{s}(k)=P_{1}a_{s}^{\frac{\alpha}{2}}$
for all $k$'s and Lemma~\ref{Lem5}. Therefore, we have \begin{equation}
R_{p}(i)>\frac{1}{25}\log\left(1+\frac{P_{0}(\sqrt{5})^{-\alpha}}{N_{0}+I_{p}+I_{sp}}\right)=K_{1}>0,\label{Eq:PrimaryRate}\end{equation}
 where the relationship that $\|X_{p,\textrm{tx}}(i)-X_{p,\textrm{rx}}(i)\|\leq\sqrt{5a_{p}}$
is used (see Fig.~\ref{Fig5}). This completes the proof. 
\end{proof}
\begin{lemma}
\label{lem:PrimaryDP}For $a_{p}(n)=k_{1}\log n/n$, the number of
primary S-D paths (including both HDPs and VDPs) that pass through
or originate from each primary cell is $O\left(n\sqrt{a_{p}(n)}\right)$
\emph{w.h.p.}. 
\end{lemma}
\begin{proof}
See the proof of Lemma 3 in~\cite{Scaling:Jeon} or the proof of
Lemma 2 in~\cite{Delay:Toumpis}.
\end{proof}
~~~~Now we give the proof for Theorem~\ref{Th1}. 

\begin{proof}
Consider the proof of the per-node throughput in~(\ref{Eq9}). According
to the definitions in Section II, we need to show that there are deterministic
constants $c_{2}>0$ and $c_{1}<+\infty$ to satisfy \begin{equation}
\lim_{n\to\infty}p\left(\frac{c_{2}}{\sqrt{n\log n}}\leq\lambda_{p}(n)\leq\frac{c_{1}}{\sqrt{n\log n}}\right)=1.\label{Eq11}\end{equation}

A loose upper bound of the per-node throughput for the primary network
is achieved when the secondary network is absent. Gupta and Kumar~\cite{Capacity:Gupta}
have already showed that such an upper bound given in~(\ref{Eq11})
exists. We then only need to consider the proof for the lower bound.

Since a given TX node in each primary cell can support a constant
data rate of $K_{1}$ (see Lemma~\ref{Lem4}), each primary S-D pair
can achieve a data rate of at least $K_{1}$ divided by the maximum
number of data paths that pass through and originate from the primary
cell. From Lemma~\ref{lem:PrimaryDP}, we know that the number of
data paths that pass through or originate from each primary cell is
$O\left(n\sqrt{a_{p}(n)}\right)$ \emph{w.h.p..} Therefore,\emph{
}the throughput per S-D pair $\lambda_{p}(n)$ is lower-bounded by
$\Omega\left(\tfrac{1}{n\sqrt{a_{p}(n)}}\right)$ \emph{w.h.p.}, i.e.,
the lower bound is $\Omega\left(\frac{1}{\sqrt{n\log n}}\right)$
$w.h.p.$. 

From Lemma~\ref{Lem2}, the number of primary S-D pairs is lower-bounded
by $\frac{n}{4}$ \emph{w.h.p.}. Thus, the sum throughput $S_{p}(n)$
is lower-bounded by $\frac{n}{4}\lambda_{p}(n)$ \emph{w.h.p.}, i.e.,
the lower bound is $\Omega\left(\sqrt{\frac{n}{\log n}}\right)$ \emph{w.h.p.}.
The upper bound of $S_{p}(n)$ is already established in~\cite{Capacity:Gupta}.
This completes the proof. 
\end{proof}
From the proof of Theorem~\ref{Th1}, the throughput per S-D pair
for the primary network can be written as

\begin{equation}
\lambda_{p}(n)=\Theta\left(\frac{1}{n\sqrt{a_{p}(n)}}\right),\quad w.h.p..\label{eq:PThroughput}\end{equation}

\subsection{Delay-throughput Tradeoff for the Primary Network}

Combining the results in~(\ref{eq:P_Delay}) and (\ref{eq:PThroughput}),
the delay-throughput tradeoff for the primary network is given by
the following theorem.

\begin{thm}
\label{thm:delay_p }With the primary protocol defined in Section
III, the delay-throughput tradeoff is

\begin{equation}
D_{p}(n)=\Theta\left(n\lambda_{p}(n)\right),\:\:\textrm{for }\;\lambda_{p}(n)=O\left(\frac{1}{\sqrt{n\log n}}\right).\label{eq:delay_p}\end{equation}

\end{thm}

\section{Delay and Throughput Analysis for the Secondary Network }

The difference between the primary and the secondary transmission
schemes arises from the presence of the preservation regions. When
their paths are blocked by the preservation regions, the secondary
relay nodes buffer the packets and wait until the next hop is available.
Due to the presence of the preservation region, the secondary packets
will experience a larger delay compared with the primary packets.
However, the average packet delay per hop for each secondary S-D data
path is still a constant as we discussed later. Thus, we can show
that the throughput scaling law and the delay-throughput tradeoff
for the secondary network are the same as those in the primary network.
In the following discussion, we first analyze the average packet delay,
then discuss the throughput scaling law, and finally describe the
delay-throughput tradeoff.

\subsection{Delay Analysis for the Secondary Network}

The average packet delay for the secondary network is given by the
following theorem.

\begin{thm}
\label{thm:SDelay}According to the proposed secondary network protocol
in Section III, the packet delay is given by

\begin{equation}
D_{s}(m)=\Theta\left(\frac{1}{\sqrt{a_{s}(m)}}\right),\quad w.h.p..\label{eq:S_Delay}\end{equation}

\end{thm}
~~~~Before giving the proof of Theorem~\ref{thm:SDelay}, we
present the following lemma.

\begin{lemma}
\label{lem:HopDelaySR}The average packet delay for each secondary
hop is $\Theta(1)$. 
\end{lemma}
\begin{proof}
Let $D_{s,h}^{j}(i)$ denote the packet delay for the secondary network
over hop $j$ and S-D pair $i$. As shown in Fig.~\ref{Fig6}, if
there are no preservation regions, each secondary cell has one active
time slot in each primary time slot. In another word, each secondary
packet will experience a worst-case delay of $t_{p}$ at each hop,
i.e., $D_{s,h}^{j}(i)=t_{p}$. When we have the preservation regions,
according to Lemma~\ref{Lem:Active}, $D_{s,h}^{j}(i)$ is a bounded
random variable. It depends on the location of the active TX from
which the secondary packet departs. As shown in Fig.~\ref{Fig6}
and Fig.~\ref{Fig:Cluster}, when the active TX is located in the
worst places as shown in Fig.~\ref{Fig:Cluster}, $D_{s,h}^{j}(i)$
is $\frac{1}{\eta_{\textrm{min}}}t_{p}$, where $\eta_{\textrm{min}}=\frac{9}{25}$
is the minimum value of the opportunistic factor $\eta$. Similarly,
when the active TX is located in the best places as shown in Fig.~\ref{Fig:Best},
$D_{s,h}^{j}(i)$ is $\frac{1}{\eta_{\textrm{max}}}t_{p}$, where
$\eta_{max}=\frac{16}{25}$ is the maximum value of the opportunistic
factor $\eta$. Hence, the ensemble average of $D_{s,h}^{j}(i)$ will
be a constant $c_{0}$, where $\frac{1}{\eta_{\textrm{max}}}t_{p}<c_{0}<\frac{1}{\eta_{\textrm{min}}}t_{p}$,
i.e., $E\left[D_{s,h}^{j}(i)\right]=\Theta(1)$. This completes the
proof.
\end{proof}
~~~~Now, let us prove Theorem~\ref{thm:SDelay}.

\begin{proof}
Since each secondary hop covers a distance of $\Theta\left(\sqrt{a_{s}(m)}\right)$
\emph{w.h.p}., and similarly as in the proof of Theorem~2, the average
length of each secondary S-D data path is $\Theta(1)$, the average
number of hops for each secondary packet is $\Theta\left(\tfrac{1}{\sqrt{a_{s}(m)}}\right)$
\emph{w.h.p}.. From Lemma~\ref{lem:HopDelaySR}, the average packet
delay for each secondary hop is $\Theta(1)$. Therefore, the average
packet delay for the secondary network is

\[
D_{s}(m)=\Theta\left(\frac{1}{\sqrt{a_{s}(m)}}\right)\]
\emph{w.h.p}., which completes the proof.
\end{proof}

\subsection{Throughput Analysis for the Secondary Network}

For the secondary network, the throughput scaling law is given by
the following theorem.

\begin{thm}
\label{Th2} With the secondary protocol defined in Section III, the
secondary network can achieve the following throughput per-node and
sum throughput \emph{w.h.p.}: \begin{equation}
\lambda_{s}(m)=\Theta\left(\sqrt{\frac{1}{m\log m}}\right)\label{Eq12}\end{equation}
 and \begin{equation}
T_{s}(m)=\Theta\left(\sqrt{\frac{m}{\log m}}\right).\label{Eq13}\end{equation}

Similarly as in the primary network case, we first present two lemmas,
then use these lemmas to prove Theorem~\ref{Th2}.
\end{thm}
\begin{lemma}
\label{lem:SR_Rate}With the proposed secondary protocol, each TX
node in a secondary cell can support a data rate of $K_{2}$, where
$K_{2}>0$ is independent of $m$. 
\end{lemma}
\begin{proof}
Due to the presence of the preservation regions, a minimum distance
of $1.5\sqrt{a_{p}}$ from all primary TXs to a specific active secondary
RX can be guaranteed. At a given secondary packet slot and at the
$i$-th secondary link (i.e., the active transmission initiated in
the $i$-th secondary cell), the interference from all active primary
TXs is upper-bounded as\begin{eqnarray}
I_{ps}(i) & < & P_{0}a_{p}^{\frac{\alpha}{2}}\sum_{t=1}^{\infty}8t((3t-1)\sqrt{a_{p}})^{-\alpha}\nonumber \\
 &  & +P_{0}a_{p}^{\frac{\alpha}{2}}\left(1.5\sqrt{a_{p}}\right)^{-\alpha}\nonumber \\
 & < & P_{0}\sum_{t=1}^{\infty}8t(3t-1)^{-\alpha}+P_{0}(1.5)^{-\alpha}\nonumber \\
 &  & =I_{ps}<\infty,\label{eq:EqIps}\end{eqnarray}
where we applied the same technique as in the proof of Lemma~\ref{Lem4}
to obtain the upper bound. Likewise, $I_{s}(i)$ is upper-bounded
by $I_{s}=P_{1}\sum_{t=1}^{\infty}8t(4t-1)^{-\alpha}$, which converges
to a constant as shown in Lemma~\ref{Lem5} (see the Appendix). Considering
the effects of the preservation region, the lower bound of the data
rate that is supported in each secondary cell can be written as\begin{eqnarray}
R_{s}(i)>\frac{1}{25}\eta_{\textrm{min}}\log\left(1+\frac{P_{0}(\sqrt{5})^{-\alpha}}{N_{0}+I_{ps}+I_{s}}\right)=K_{2}>0,\end{eqnarray}
where $\eta_{\textrm{min}}=\frac{9}{25}$ represents the penalty due
to the presence of the preservation region. Thus, we can guarantee
a constant data rate $K_{2}>0$ for a given TX node in each secondary
cell, which completes the proof. 
\end{proof}
\begin{lemma}
\label{lem:SecondaryDP}For $a_{s}(m)=k_{2}\log m/m$, the number
of secondary S-D paths (including both HDPs and VDPs) that pass through
or originate from each secondary cell is $O\left(m\sqrt{a_{s}(m)}\right)$
\emph{w.h.p.}. 
\end{lemma}
\begin{proof}
The proof of Lemma~\ref{lem:SecondaryDP} follows the same logic
as that in the proof of Lemma~\ref{lem:PrimaryDP}.
\end{proof}
~~~~Now, let us prove Theorem~\ref{Th2}.

\begin{proof}
The proof of Theorem~\ref{Th2} is similar to the proof of Theorem~\ref{Th1}. 
\end{proof}
~~Similarly as in Theorem~2, the throughput per S-D pair of the
secondary network can be written as

\begin{equation}
\lambda_{s}(m)=\Theta\left(\frac{1}{m\sqrt{a_{s}(m)}}\right),\quad w.h.p..\label{eq:S_Throughput}\end{equation}

\subsection{Delay-throughput Tradeoff for the Secondary Network}

Combining the results in~(\ref{eq:S_Delay}) and (\ref{eq:S_Throughput}),
the delay-throughput tradeoff for the secondary network is given by
the following theorem.

\begin{thm}
\label{thm:delay_s}With the secondary protocol defined in Section
III, the delay-throughput tradeoff for the secondary network is

\begin{equation}
D_{s}(m)=\Theta\left(m\lambda_{s}(m)\right),\:\:\textrm{for }\lambda_{s}(m)=O\left(\frac{1}{\sqrt{m\log m}}\right).\label{eq:delay_s}\end{equation}

\end{thm}

\section{Conclusion}

In this paper, we studied the coexistence of two wireless networks
with different priorities, where the primary network has a higher
priority to access the spectrum, and the secondary network opportunistically
explore the spectrum. When the secondary network has a higher density,
with our proposed protocols, both of these networks can achieve the
throughput scaling law promised by Gupta and Kumar in~\cite{Capacity:Gupta}.
Comparing with the recent result in~\cite{Scaling:Jeon}, we only
assumed the knowledge about the primary TX locations and there is
no outage penalty for the secondary nodes. By using a fluid model,
we also showed that both networks can achieve the same delay-throughput
tradeoff as the optimal one established for a stand-alone wireless
network in~\cite{Delay:Gamal1}.

\appendix{In the appendix, we first recall the useful Chernoff bound for a
Poisson random variable from~\cite{Random:Mitzenmacher}; then we
give a lemma to show that the infinite series sums in Lemma~\ref{Lem4}
and Lemma~\ref{lem:SR_Rate} converge to a constant.}

\begin{lemma}
\emph{\label{lem:Mitz}(Theorem 5.4 in~\cite{Random:Mitzenmacher})
}Let $X$ be a Poisson random variable with parameter $\mu$.
\end{lemma}
\begin{enumerate}
\item If $x>\mu$, then\[
p\left(X\geq x\right)\leq\frac{e^{-\mu}(e\mu)^{x}}{x^{x}};\]

\item If $x<\mu$, then\[
p\left(X\leq x\right)\leq\frac{e^{-\mu}(e\mu)^{x}}{x^{x}}.\]

\end{enumerate}
\begin{lemma}
\label{Lem5}The sum $\sum_{t=1}^{\infty}at(bt-1)^{-\alpha}$ converges
to a constant, where $\alpha>2$, $a$ and $b$ are positive integers.
\end{lemma}
\begin{proof}
\begin{eqnarray}
\sum_{t=1}^{\infty}\frac{at}{(bt-1)^{\alpha}} & = & \frac{a}{b^{\alpha}}\sum_{t=1}^{\infty}\frac{t}{(t-\frac{1}{b})^{\alpha}}\nonumber \\
 & = & \frac{a}{b^{\alpha}}\sum_{t=1}^{\infty}\frac{1}{(t-\frac{1}{b})^{\alpha-1}}\nonumber \\
 &  & +\frac{a}{b^{\alpha+1}}\sum_{t=1}^{\infty}\frac{1}{(t-\frac{1}{b})^{\alpha}}.\label{Eq8}\end{eqnarray}
Applying the following inequality

\[
\sum_{t=1}^{\infty}\frac{1}{(t-\frac{1}{b})^{\alpha}}\leq\frac{1}{(1-\frac{1}{b})^{\alpha}}+\int_{1}^{\infty}\frac{1}{(t-\frac{1}{b})^{\alpha}}\, dt\]
to~(\ref{Eq8}), we obtain\begin{eqnarray*}
\sum_{t=1}^{\infty}\frac{at}{(bt-1)^{\alpha}} & \leq & \frac{a}{b^{\alpha}}\left(\frac{1}{(1-\frac{1}{b})^{\alpha-1}}+\int_{1}^{\infty}\frac{1}{(t-\frac{1}{b})^{\alpha-1}}\, dt\right)\\
 &  & +\frac{a}{b^{\alpha+1}}\left(\frac{1}{(1-\frac{1}{b})^{\alpha}}+\int_{1}^{\infty}\frac{1}{(t-\frac{1}{b})^{\alpha}}\, dt\right)\\
 & = & \frac{a}{b^{\alpha}\left(1-\frac{1}{b}\right)^{\alpha-1}}+\frac{a\left(1-\frac{1}{b}\right)^{-\alpha+2}}{b^{\alpha}(\alpha-2)}\\
 &  & +\frac{a}{b^{\alpha+1}\left(1-\frac{1}{b}\right)^{\alpha}}+\frac{a\left(1-\frac{1}{b}\right)^{-\alpha+1}}{b^{\alpha+1}(\alpha-1)},\end{eqnarray*}
where the last equation is a constant when $\alpha>2$. This completes
the proof.
\end{proof}

\section*{Acknowledgment}

The authors would like to thank Dr. Tie Liu for helpful discussions
during the course of this work.

\end{document}